\documentclass{article}
\usepackage{graphicx}
\usepackage{amsmath,amssymb}
\usepackage{epstopdf}
\usepackage{epsfig}
\usepackage{mathrsfs}
\usepackage[breaklinks=true]{hyperref}

\begin{document}
\title{Before thy throne I slip in gently, hoping nobody will notice: A memorial to Robin Hudson (1940-2021)}

\author{John E.~Gough \\
\texttt{jug@aber.ac.uk}\\
Aberystwyth University, SY23 3BZ, Wales, United Kingdom}
 
\date{\today}

\maketitle

\begin{abstract}
We dedicate this to the life and work of Robin Hudson - a mathematical physicist who developed the peerless quantum stochastic calculus, but who also inspired generations of researchers with both his intellect and wit.
\end{abstract}

\section{Introduction}
Robin Lyth Hudson was an English mathematician who made fundamental contributions to mathematical physics, foremost amongst these being the construction of quantum stochastic calculus with K.R. Parthasarathy. He was a constant presence at QP conferences over the years, often offering profound insights on topics. Despite his large barrel-chested frame and his formidable research standing, Robin's natural instinct was always to avoid taking centre stage, making him both unobtrusive yet highly approachable at the same time. His passions were simple: family, mathematics, music (especially Bach) and fairness. His sense of fairness was uncompromising, and his ferocious honesty was the platform for his brilliant sense of humour.

The following is a personal and inevitably flawed description of Robin Hudson. Rather than aiming for a definitive, or even accurate account, I've plumbed for something that captures a sense of his spirit and imagination. 

I first met Robin Hudson in March 1994 at a workshop in Oberwolfach organized by Luigi Accardi and Wilhelm von Waldenfels. His famous paper \cite{HP84} with K.R. Parthasarathy (Partha) was reaching its tenth birthday at that time and had already made a considerable impact in the mathematical physics community as the canonical way to extend the Ito calculus to the quantum domain. I had just started on a postdoc with Luigi and was still new to the field of quantum probability (QP), but had read the paper the previous year and was looking at quantum stochastic models arising as weak coupling limits. The following June I met both Robin and Partha together at a workshop (the IXth QP conference) at Marseille. Later, I shared an office with Robin when working at Nottingham Trent, and many of the (unsubstantiable) comments I present stem from this time. The comments in italics in this text come from Robin's fragmentary autobiographical notes edited by his wife Olga, and should be taken with substantially more credence by the reader than what I claim to have heard from Robin in our office on the now-demolished site of the Goldsmith building in Nottingham.

\begin{figure}
	\centering
		\includegraphics[width=0.750\textwidth]{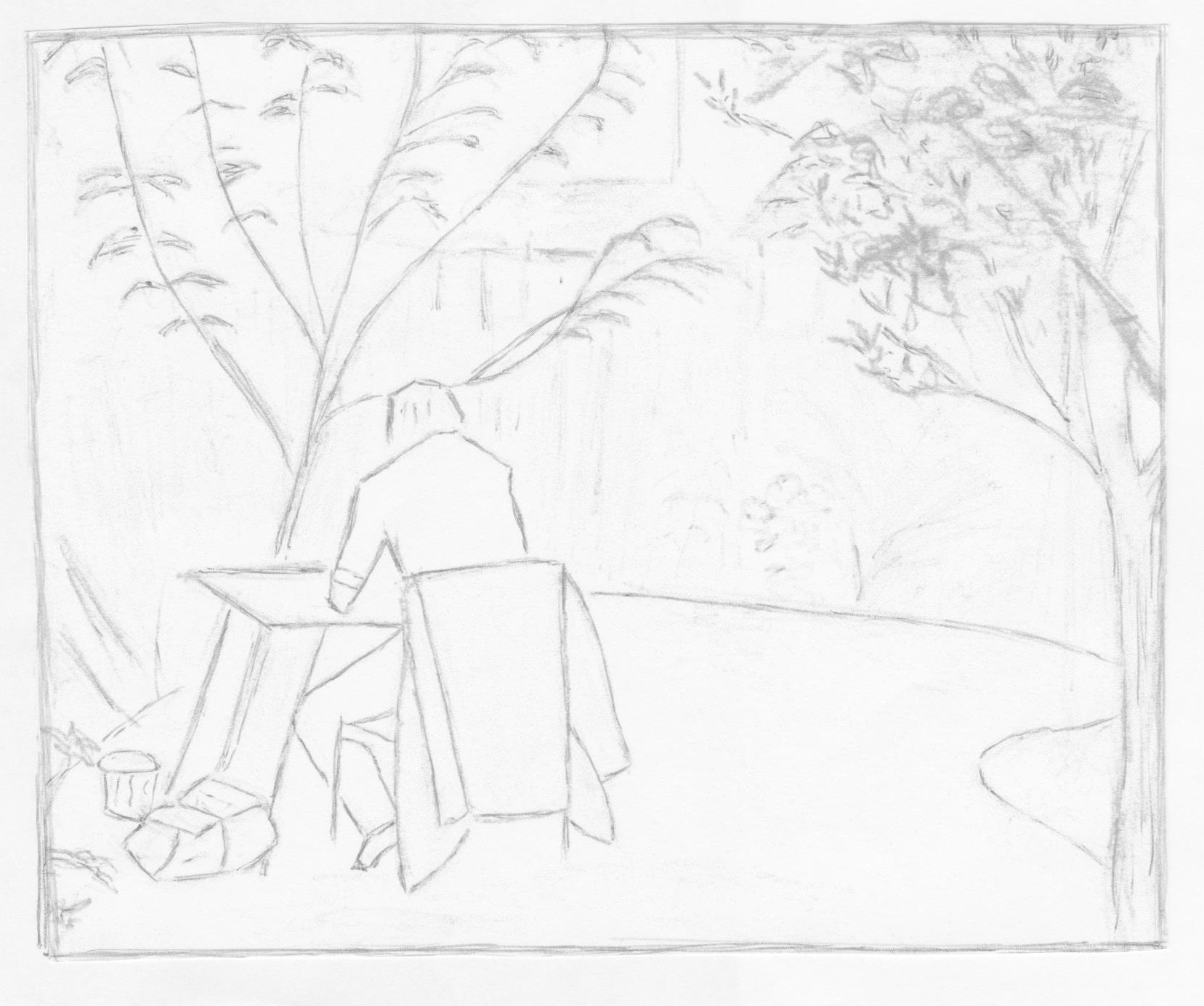}
	\caption{Robin at work on a maths problem in his Nottingham back garden in 1977. (Courtesy of Olga Hudson)}
	\label{fig:RobinWorking}
\end{figure}

Robin was born on May 4th 1940 in, of all places, Aberdeen. \lq\lq\textit{My parents were not Scots , so my birth in Aberdeenshire ... could be regarded as fortuitous. My father came from a long line of ministers of religion in the Methodist Church on both paternal and maternal sides.}\rq\rq In fact, Robin's father - Frederick \lq\lq Will" Lyth Hudson - was a research chemist: he obtained an undergraduate degree at Manchester and a doctorate at Berkeley (California) before returning to Britain during the Great Depression to become a paper mill chemist for the paper manufacturers Wiggins Teape. Both parents were Quakers at the time and enrolled (the Quaker analogue of baptism) their children as well. Will had moved to Scotland before WWII and was Clerk of the Aberdeen Friends (Quaker) Meeting which covered much of Eastern Scotland. 
At the time, being quasi-teetotal pacifists, and Sassenachs to boot, did not endear them to locals. But he returned to England in 1945 to work at BIP (formerly British Cyanides Company) in Birmingham and here developed plastic coated bags. About a decade later, he switched to an academic career becoming the head of the Paper Science Department at Manchester (UMIST). Robin’s mother was a poet; something which Olga feels, surely also influenced the mathematics he most admired; research earthed in physical reality but written with simplicity and elegance: a fusion of truth and beauty.

Robin left for England (Kidderminster) circa 1945. \lq\lq\textit{Of my early life in Aberdeenshire I also remember surprisingly little .. the pleasure of retrieving our hens’ eggs from the henhouse and the puzzling sight of the Ford 8 car, forlorn, silent and unmoving in the garage… In due course my meagre wartime collection of toys were packed into a trunk and entrusted to the London and North Eastern Railway, which lost them!}\rq\rq 

On moving South the Hudson household began to attend the Stourbridge Friends (Quakers) Meeting.
As Robin recalled \lq\lq \textit{I got through the [weekly] hour of silence and enforced inactivity by observing on the clock when half the time had passed at half past ten, then two thirds at twenty to eleven, then three quarters at quarter to, then four fifths at twelve minutes to, five sixths at ten to, and so on. From six sevenths onwards, as well as minutes, seconds and even fractions of seconds were involved which became increasingly difficult to calculate. But the joy of this mind game was that the successive times came faster and faster than I could calculate them until that glorious release when the two most senior attendees shook hands and it was all over till next Sunday. It would be fanciful to say that I acquired thereby an interest in the mathematical concept of limit.}\rq\rq

Robin was given the opportunity to formally declare if he wished to remain a Quaker when he was 16. (This is the Quaker analogue of the sacrament of Confirmation, though allowing enrollees to be older and maturer when making their decision.) At any rate, Robin decided to decline.

At the local Stourbridge Grammar School, Robin discovered a passion for music that would last a lifetime - with frequent interruptions from mathematical concerns. \lq\lq\textit{I also started composing, into a blank musical manuscript book whose cover helpfully listed all twelve bass and treble clef keys signatures. I became fascinated with questions such as why two sharps indicated B minor or D major, both keys having white key tonics, whereas two flats indicated G minor or B flat major, the latter having a black key tonic, which seemed to me to break all symmetry between sharps and flats [further elaborations of this] ... It was many years before I realised that mathematics might offer both a descriptive language of and an explanation for these patterns.}\rq\rq

\lq\lq\textit{Somehow the key of B minor itself struck some resonance, acquiring a new resonance in me which is difficult to explain. But one of my first LPs was of two of the Bach orchestral suites, including the B minor one. I think I was initially attracted to the well known portrait picture of Johann Sebastian adorning the sleeve ... Another much loved early possession was a Deutsche Gramophon compilation of contrapuntal music called \lq\lq Musik[k]unde in Beispielen\rq\rq\,  [ which had nothing in B minor] ... Bach was, of course, well represented ...  But the excerpt which immediately gripped me and does to this day, still sending the proverbial shivers down my spine, was a movement from the unaccompanied choral motet \lq\lq Jesu, meine Freude”… I was enthralled by the three-part setting of the verse “Gute Nacht, O Wiesen”. Before the first line of the chorale proper, the rather prosaic words \lq\lq Gute Nacht” are sung three times, on the third hearing the alto line repeats the three syllables on the painful but at first consoling sixth note of the minor scale. Meanwhile the tenor voice descends to achieve the harmonic interval of a diminished seventh on the third syllable. My hairs stood on end! But was it Bach’s genius or the angelic boys’ voices of the Leipzig Thomaserchor that so moved me? Since the enchantment persists it must be the former. But for some years it was not so clear.}\rq\rq 

\bigskip

Robin became a doctoral student at Oxford University in the Quantum Chemistry group headed by Charles Coulson, under the supervision of John Lewis. He was, in fact, John’s first doctoral student. At that stage, mathematical physics was still looked upon with a degree of suspicion, in particular, John's insistence that functional analysis had any role to play in physics. \lq\lq \textit{John Trevor Lewis, JTL, was born a Welshman in Swansea, but moved to Belfast at the age of 15, and became passionate about his new Irish heritage. After completing his doctorate in Belfast he had a research position in Coulson’s group at Oxford.}\rq\rq

Robin's PhD/DPhil experience was not so straightforward: \lq\lq \textit{John had become a fellow of Brasenose College with a University Lectureship. His method of supervision was to suggest various potential interesting research papers for me to study ... and to hope that I would find a suitable research topic for myself. This was quite a hard initiation and it is perhaps unsurprising that it was modified for his subsequent doctoral students. After struggling to understand many papers, I nevertheless found a problem, the group theory of the anisotropic multidimensional quantum mechanical harmonic oscillator, which I could understand and eventually master. A distinguished visitor to Coulson’s group, the Swiss mathematical physicist and author of a well known book on foundations of quantum theory J M Jauch, who had worked on this topic was interested and complimentary when I discussed it with him. ... In my third and final postgraduate year I discovered that a Russian named Demkov had also studied my problem and had published his results, similar to mine, in a respectable Soviet journal. I concluded that my work would not earn me a DPhil and looked instead for another problem. What eventually materialised was a topic so esoteric as to make duplication very unlikely. But nearly all the work on this was carried out in the first two years of my first academic appointment at Nottingham, a fact which must arouse the envy of those who nowadays struggle through doctoral studies and then a succession of post-doctoral research positions in search of a proper academic position.}\rq\rq

George Mackey visited Oxford and gave a series of lectures on systems of imprimitivity and group representations. Robin took on the major task of writing the notes. This would be an introduction to the canonical commutation relations from the point of view of  Weyl and Stone-von Neumann. For a clear presentation on Mackey's ideas and their impact on quantum field theory (as well as quantum stochastic calculus), see the article by Varadarajan \cite{Varadarajan}.

Robin eventually received his doctorate from Oxford in 1966 under John T. Lewis with a thesis entitled \textit{Generalised Translation-Invariant Mechanics}.

\bigskip

On the subject of career development, Robin offered the following: \lq\lq\textit{In an academic world spanning 55 years, based on several institutions, I never had a successful job interview resulting in an appointment. My first attempt was at the University of Sussex at Falmer outside Brighton. I was brought down by an unwise attempt to disguise my distaste for computers and their increasing role in mathematics teaching and research. Next, it was off to Liverpool at Coulson’s urging. But John told me not to touch it. Then I tried for a junior research fellowship at Jesus College, Oxford. All seemed to go well until the most important part, the ceremony of drinking port after dinner. The port was laid down, it was carefully explained, thirty years previously. I fumbled with the glass and overturned it. Thereafter I relaxed, nothing to lose, and enjoyed a selection of malt whiskies.}\rq\rq

At this stage, fate intervened to force a move to Nottingham. \lq\lq\textit{Coulson recommended Nottingham probably because George Hall, newly appointed professor of applied mathematics there, was also a quantum chemist ... This time John agreed, perhaps because George was an Ulsterman, though of the opposite persuasion in the great religious divide, his father having had a \lq\lq long and glorious ministry” in the service of the Presbyterian Church. ... I did not go to Nottingham for interview, Nottingham came to me. Having obtained permission from the university Senate for this unusual procedure, George turned up in Oxford with full powers to appoint me to an assistant lectureship at an annual salary of £1050, £50 above the minimum.}"

Robin once told me a story of his time chairing quantum chemistry seminars, a subject so boring that he survived only by mastering how to sleep with his eyes open and wake up after 55 minutes, just in time to do the formalities of thanking the speaker and asking the audience for questions. This procedure worked fine until one guest speaker finished his talk a quarter of an hour earlier than scheduled, after which there was an awkward few minutes before Robin emerged from trance \footnote{I am grateful to Olga Hudson for not only confirming this, but adding that he continued to do this for a great many years afterwards during boring talks - evidently with greater success than we may have noticed!}. 

In 1982, the first conference in Quantum Probability was organized by Luigi Accardi at Villa Mondragone, see \cite{QPI}. The conference gathered together many of the key players - Robin, Partha, Luigi, Wilhelm von Waldenfels, Alexander Holevo, Alberto Frigerio, Vittorio Gorini, etc. - for the first time in one location. This was to become an enduring series of workshops which established the field internationally and which led to several volumes of Lecture Notes in Mathematics, and Robin was to be a regular attender. 

He was promoted to Chair in 1985 and served as Head of the Mathematics Department from 1987 to 1990. This was followed by positions at Nottingham Trent and Loughborough University.

Olga adds: In his last years, a time of increasing pain, Robin felt greatly sustained by the mathematical community. In 2013, he was surprised and delighted by the award of his honorary doctorate from \L odz University and subsequently in the following year by John Gough’s revelatory sequence of quantum probability conferences in Cambridge. Then, in Summer 2016, when he was no longer able to travel abroad, Robin had the unexpected pleasure of meeting many of his international colleagues (from both the QP and the quantum structures communities) for the last time at two consecutive conferences at Leicester University. His last effort, in the long Covid summer of 2020, was to address a Zoom seminar from Portsmouth University: a cause of considerable technological anxiety, were it not for the gentle pre-seminar coaching of Michael Gnacik.

In 2013 he became a doctor \textit{honoris causa} of  \L odz University.

Robin gave a number of general lectures about mathematics and music. \lq\lq \textit{I think that the affinity [of mathematics] to Bach's music has more to do with complexity, with tightness and discipline of structure, and above all with a sense that what is proclaimed repeatedly in the fugue is true. The best fugues also spring surprises, but they are surprises which are seen to be inevitable in retrospect. Mathematics is a bit like that; the mathematician van der Pohl said that the best mathematics is both surprising and inevitable. But the only answer I can give is that it is a mystery.}"

\section{Robin's Work}
The following section is based around Robin's opening presentation to the 2014 workshop on Quantum Control at the Newton Institute in Cambridge.
More details can be found in Robin's papers \cite{Hudson_Early}, \cite{Hudson_walk}

Heisenberg's canonical equations take the form $\left[ q,p\right] =i\hbar $ where Planck's constant is $\hbar =1.05457\times 10^{-34}$ Nms which, as Robin was wont to say, is \lq\lq\textit{quite small in everyday units}.\rq\rq    We will use units for which $\hbar =2$, that is, 
\begin{eqnarray}
\left[ q,p\right] =2i.
\label{eq:CCR}
\end{eqnarray}
As is well known, the Schrodinger representation of these operators is $\left( q\psi \right) \left( x\right) =x\psi \left( x\right) $ and $\left(p\psi \right) \left( x\right) =-i2\partial \psi \left( x\right) $, and a key mathematical result is the Stone-von Neumann Theorem which states that every irreducible representation is unitarily equivalent to the Schrodinger one. The annihilation operator $a$ and its adjoint, the creator $a^\ast$, are defined via
\begin{eqnarray}
q= a+a^\ast , \qquad p = \frac{1}{i} ( a - a^\ast ) ,
\end{eqnarray}
and satisfy the commutation relations $[a,a^\ast ] =1$. The number operator $N = a^\ast a$ has, as is well-known, the non-negative integers as eigenvalues with \textit{ground state} $|0 \rangle$ being the eigenvector for eigenvalue $n=0$. The observables $q$ and $p$ then both have standard Gaussian (i.e., mean zero and unit variance) distribution in this state: this incidentally is the reason for the strange choice $\hbar =2$ in (\ref{eq:CCR}).

Let $\langle \cdot \rangle $ be a quantum state, then the characteristic
function is 
\begin{eqnarray}
\langle e^{iuq+pv}\rangle =tr\left\{ \rho e^{iuq+ivp}\right\}
=\int_{R^{2}}e^{iux+ivy}w\left( x,y\right) dxdy
\end{eqnarray}
where $\rho $ is the associated density matrix and $w$ is the Wigner distribution function. The covariance matrix is 
\begin{eqnarray}
C=\left[ 
\begin{array}{cc}
\langle q^{2}\rangle -\langle q\rangle ^{2} & \langle qp\rangle -\langle
q\rangle \langle p\rangle  \\ 
\langle pq\rangle -\langle p\rangle \langle q\rangle  & \langle p^{2}\rangle
-\langle p\rangle ^{2}
\end{array}
\right] 
\end{eqnarray}
and there always exists a linear (symplectic) transformation which converts the matrix to the form 
\begin{eqnarray}
C=\left[ 
\begin{array}{cc}
\sigma ^{2} & i \\ 
-i & \sigma ^{2}
\end{array}
\right] .
\end{eqnarray}
with $\sigma \geq 1$ and we refer to this as the standard from.

It is instructive to look at mean zero Gaussian states. Let us define the family of states having density matrices $\rho_\theta = \frac{1}{Z(\theta )} e^{-\theta N}$ for $\theta \ge 0$ with the normalization is $Z (\theta ) = (1-e^{-\theta   } )^{-1}$. The expectation of the number operator is given by $\bar n (\theta ) = (e^\theta -1 )^{-1}$. These states are all mixed and faithful (that is, $\rho_\theta$ has no zero eigenvalues) however the limit $\theta \to \infty$ yields the pure state corresponding to $| 0 \rangle$. For the family, we find a Gaussian characteristic function is 
\begin{eqnarray}
\langle e^{iuq+pv}\rangle =e^{-\frac{1}{2}\sigma ^{2}(u^{2}+v^{2})}
\end{eqnarray}
with $ \sigma^2 = 2 \bar n + 1 \equiv \coth (\theta /2)$. We note that these states may be obtained using the Araki-Woods construction where we introduce two modes $a_1$ and $a_2$ and set
\begin{eqnarray}
a = \sqrt{ \bar n +1 } a_1 \otimes 1_2 + \sqrt{ \bar n} 1_1 \otimes a_2^\ast.
\end{eqnarray}
 with state $| 0 \rangle_1 \otimes | 0 \rangle_2$.

The Wigner distribution functions for this family (and also the ground state $\theta = \infty$) turn out to be Gaussian functions. 

\subsection{Positivity of the Wigner Distribution}
Typically however Wigner distributions will be negative and an important early result of Robin is \textit{Hudson's Theorem}  \cite{Hudson_Wigner} that a
pure state will lead to a positive Wigner distribution if and only if it takes the corresponding wavefunction form
\begin{eqnarray}
\psi \left( x\right) =e^{\frac{1}{2}ax^{2}+bx+c}
\end{eqnarray}
where $a$ has negative real part. The result was extended to $n$ dimensions by Soto and Clavarie \cite{Soto_Claverie}. We also mention an elegant alternative proof based on the properties of pseudo-differential operators \cite{Toft}.

These are coherent states. What makes Robin's work here all the more impressive is the fact that this was an unexplored area of investigation at the time, but has since emerged as the key example on non-classicality with Gaussian states considered the most classical of quantum states. However, as Robin would point out, every quantum state is non-classical - it's just a question of seeking an appropriate non-classical behaviour.

\subsection{Quantum De Moivre-Laplace Theorem}
Robin's first PhD student was Clive Cushen at Nottingham. Together they produced a paper on a quantum central limit theorem \cite{HC}.
Let $\left( q_{1},p_{1}\right) ,\left( q_{2},p_{2}\right) ,\cdots $ be a sequence of canonical pairs with $\left[ q_{j},p_{k}\right] =i\hbar \delta _{jk}$, and $\left[ q_{j},q_{k}\right] =0=[p_{j},p_{k}]$. Suppose that we have a joint state where the pairs are independent and identically distributed with mean zero, then one considers the sums 
\begin{eqnarray}
Q_{n}=\frac{1}{\sqrt{n}}\sum_{j=1}^{n}q_{j},\quad P_{n}=\frac{1}{\sqrt{n}}%
\sum_{j=1}^{n}p_{k}.
\end{eqnarray}
If the covariances are finite one might expect that there is a form of quantum central limit theorem. Some observations are in order. First, $\left( Q_{n},P_{n}\right) $ is again a canonical pair, that is, $\left[ Q_{n},P_{n}\right] =i\hbar $. Second, the result will be a genuine
generalization of the classical central limit theorem since there can be no underlying joint probability distribution. The analogue of the de Moivre-Laplace Theorem is that the Wigner distributions converge to a limit Gaussian one - specifically if the covariances are in the standard form $C$ then we obtain the Gaussian with mean zero and Gaussian $C$. The result may actually be framed in terms of the stronger sense of density matrices: $tr\{\rho _{n}$ $X\}\rightarrow tr\{\rho X\}$ for each bound $X$.

\lq\lq\textit{Clive’s PhD topic was mainly a central limit theorem for canonical pairs, that is, pairs such as p and q satisfying the Heisenberg relation. In classical probability theory the central limit theorem explains the ubiquity of the well known bell-shaped curve, that is, the Gaussian distribution, for the distribution of accumulated errors from a mean value. The classical theory deals only with commuting quantities so a new form of probability theory, now called quantum probability , was needed for its correct formulation. A joint paper with Clive was my first publication. It led on logically to most of my later scientific work and has perhaps not had as much recognition as it deserved.}\rq\rq

It is a short step then to define processes 
\begin{eqnarray}
Q_{n}\left( t\right) &=&\frac{1}{\sqrt{n}}\sum_{j=1}^{n\left( t\right) }q_{j}+%
\frac{nt-n\left( t\right) }{\sqrt{n}}q_{n\left( t\right) +1},\nonumber \\
P_{n}\left( t\right) &=& \frac{1}{\sqrt{n}}\sum_{j=1}^{n\left( t\right) }p_{k}+%
\frac{nt-n\left( t\right) }{\sqrt{n}}p_{n\left( t\right) +1},
\end{eqnarray}
where $n\left( t\right) $ is $nt$ rounded down to the nearest integer. The family of processes $\left\{ Q_{n}\left( t\right) :t\geq 0\right\} $ is commutative and as an essentially classical process converges to a Wiener process $Q$ with variance $\sigma ^{2}$ by the Donsker invariance principle \cite{Donsker}. The same is true for the momenta terms and so we end up with a pair of non-commuting processes which are separately Wiener processes. In fact, we readily see that the limit processes should satisfy 
\begin{eqnarray}
\left[ Q\left( t\right) ,P\left( s\right) \right] =2i\min \left( t,s\right) .
\end{eqnarray}
The question then arises as to how we could concretely realize these limit processes. The answer will turn out to be as operators on the Fock space over the square integrable functions of time: we discuss this in the next subsection. The discrete time versions can be embedded into this setting which allows us to go beyond convergence in distribution, and there by now is a wide literature on this topic.

Remarkably, even though the processes do not commute, they do have a degree of stochastic independence in the sense that

\begin{eqnarray}
\left\langle e^{i\sum_{j}u_{j}Q\left( t_{j}\right) +i\sum_{k}v_{k}P\left(
t_{k}\right) }\right\rangle =\left\langle e^{i\sum_{j}u_{j}Q\left(
t_{j}\right) }\right\rangle \left\langle e^{i\sum_{k}v_{k}P\left(
t_{k}\right) }\right\rangle 
\end{eqnarray}
which justifies the name \textit{quantum planar Brownian motion}.

\subsection{Stochastic Calculus For Quantum Planar Wiener Processes}
At this stage Robin's work with his main collaborator K.R. Parthasarathy (Partha) starts to take off \cite{Hudson_Early}.

\lq\lq\textit{Some time in 1972 a notice appeared on the display board of the Nottingham mathematics department for seminars and similar events at other universities. Dr K. R. Parthasarathy was to give some advanced lectures on Wigner densities, Weyl quantization and related topics at Sheffield University. A telephone number was given for would-be attenders. I was becoming interested in these topics and called the number. Partha did not like the telephone, along with other modern “gadgets”. Neither did I. The conversation went something like this:}

\textit{RLH: “Hello, is that Dr Parthasarathy?” \\
\indent Partha (loudly and slowly): “Yes”.\\
\indent RLH: ”Are you giving some lectures on Wigner densities, Weyl quantization and related topics?” \\
\indent Partha (loudly and slowly): “Yes”. I had been hoping for a friendly invitation to come along. But none was forthcoming. \\
\indent RLH: “Well, that’s alright then.” \\
\indent Partha (loudly and slowly): “Yes.” \\
So ended my first contact with Partha, who was to become my great friend and collaborator over a lifetime.}\rq\rq

\bigskip

The processes are naturally constructed in the setting of Bose C*-algebras. The $\sigma =1$ case is straightforward and the corresponds to the standard Fock space construction with the state being the Fock vacuum state. The $\sigma >1$ case is in one sense easier to work with as the state is now faithful: it is often referred to as the non-Fock case and can be realized using the double Fock spaces by the Araki-Woods construction.

\bigskip 

We recall that the Fock space $\Gamma \left( \mathfrak{h} \right) $ over a Hilbert space $\mathfrak{h}$ is defined as $\bigoplus_{n=0}^{\infty }\left( \mathfrak{h}^{\otimes n}\right) _{\text{symm.}}$ - the infinite direct sum of symmetric parts of the $n$-fold tensor products of $\mathfrak{h}$. The exponential vectors are defined by 
\begin{eqnarray}
|\exp \left( f\right) \rangle =1\oplus f\oplus \left( \frac{f\otimes f}{%
\sqrt{2!}}\right) \oplus \left( \frac{f\otimes f\otimes f}{\sqrt{3!}}\right)
\oplus \cdots 
\end{eqnarray}
for each $f\in \mathfrak{h}$. The annihilation operator with test function $g$ is then defined as
\begin{eqnarray}
a\left( g\right) |\exp \left( f\right) \rangle =\langle g|f\rangle |\exp
\left( f\right) \rangle ,
\end{eqnarray}
with the creator being
\begin{eqnarray}
a\left( g\right) ^{\ast }|\exp \left( f\right) \rangle =\left. \frac{d}{du}%
|\exp (f+ug)\rangle \right| _{u=0}.
\end{eqnarray}

The construction of quantum Wiener processes is as follows: one takes $\mathfrak{h}=L^{2}\left( \mathbb{R}_{+},dt\right) $ and define the annihilation and creation processes as
\begin{eqnarray}
A_{t}=a\left( \chi _{[ 0,t)}\right) ,\quad A_{t}^{\ast }=a\left( \chi
_{[ 0,t)}\right) ^{\ast },
\end{eqnarray}
then one sets
\begin{eqnarray}
Q_{t}=A_{t}+A_{t}^{\ast },\quad P_{t}= \frac{1}{i} \left( A_{t}-A_{t}^{\ast }\right) .
\end{eqnarray}

Quantum stochastic integrals $M\left( t\right) =\int_{0}^{t}\left( F\left( s\right) dA_{s}^{\ast }+G\left( s\right) dA_{s}+H_{s}dT\left( s\right) \right) $ are defined as (\textit{1st Fundamental Formula})

\begin{gather}
\langle \exp \left( f\right) |M_{t}|\exp \left( g\right) \rangle
\nonumber \\ =\int_{0}^{t}\langle \exp \left( f\right) |f\left( s\right) ^{\ast }F\left(
s\right)  +G\left( s\right) g\left( s\right)  +H (s) \, |\exp \left(
g\right) \rangle ds.
\end{gather}

An essential feature of Fock spaces is that $\Gamma \left( \mathfrak{h}_{1}\oplus \mathfrak{h}_{2}\right) \cong \Gamma \left( \mathfrak{h}_{1}\right) \otimes \Gamma \left( \mathfrak{h}_{2}\right) $ and this leads to the continuous tensor product decomposition
\begin{eqnarray}
\Gamma \left( L^{2}\left( \mathbb{R}_{+},dt\right) \right) \cong \Gamma \left(
L^{2}\left( [0,t),\right) dt\right) \otimes \Gamma \left( L^{2}([t,\infty
),dt)\right) 
\end{eqnarray}
which separates for each time $t>0$ the Fock space into a past factor and a future factor.

Let us take the integrands $F,G,H$ of the integral $M_{t}$ to be adapted - that is, they are trivial on the future factors for each $t$. The integral can be understood as a sum for adapted integrands which are step functions and the general form follows in the usual manner by approximation. The
\textit{2nd Fundamental Formula} is 
\begin{gather}
\langle M_{1}\left( t\right) \exp \left( f\right) |M_{2}(t)\exp \left(
g\right) \rangle  \nonumber \\
\int_{0}^{t}\langle \left( g\left( s\right) ^{\ast
}F_{1}\left( s\right) +G_{1}\left( s\right) f\left( s\right)
+H_{1}\left( s\right) \right) \exp \left( f\right) |M_{2}\left(
s\right) \exp \left( g\right) \rangle ds \nonumber \\
+\int_{0}^{t}\langle M_{1}\left( s\right) \exp \left( f\right) |\left(
f\left( s\right) ^{\ast }F_{2}\left( s\right) +G_{2}\left( s\right)
g\left( s\right) +H_{2}\left( s\right) \right) \exp \left( g\right)
\rangle ds \nonumber \\
+\int_{0}^{t}\langle F_{1}\left( s\right) \exp \left( f\right)
|F_{2}\left( s\right) \exp \left( g\right) \rangle 
\end{gather}
In differential form, this becomes (\textit{the quantum Ito formula})
\begin{eqnarray}
d\left( XY\right) =\left( dX\right) Y+X\left( dY\right) +\left( dX\right)
\left( dY\right) 
\end{eqnarray}
with the last term being calculated from the quantum Ito table
\begin{eqnarray}
\begin{tabular}{l|lll}
$\times $ & $dA$ & $dA^{\ast }$ & $dT$ \\ \hline
$dA$ & 0 & $dT$ & 0 \\ 
$dA^{\ast }$ & 0 & 0 & 0 \\ 
$dT$ & 0 & 0 & 0
\end{tabular}.
\end{eqnarray}

One may realize the non-Fock possibility relatively easily using a pair of Fock spaces, the joint Fock vacua as states, and the Araki-Woods construction with
\begin{eqnarray}
A_t = \sqrt{ \bar n +1 } A_1 (t) \otimes 1_2 + \sqrt{ \bar n} 1_1 \otimes A_2^\ast (t) .
\end{eqnarray}
Here the the quantum Ito table becomes
\begin{eqnarray}
\begin{tabular}{l|lll}
$\times $ & $dA$ & $dA^{\ast }$ & $ dT$ \\ \hline
$dA$ & 0 & $ ( \bar n+1 ) \, dT$ & 0 \\ 
$dA^{\ast }$ &  $  \bar n \,dT$ & 0 & 0 \\ 
$dT$ &  & 0 & 0
\end{tabular}.
\end{eqnarray}

\subsection{The Gauge Process}
Let $U$ be a unitary operator on $\mathfrak{h}$, then a unitary operator $\Gamma (U)$ is defined on the Fock space $\Gamma (\mathfrak{h})$ by $\Gamma (U) | \exp (f) \rangle = | \exp( Uf ) \rangle$. In particular, a self-adjoint operator $M$ generates a strongly continuous group of unitary operators on $\mathfrak{h}$ which in turn extends to a strongly continuous group of unitaries on the Fock space with Stone generator $d\Gamma (M)$ known as the differential second quantization of $M$. 

The gauge process $\Lambda (t)$ is the differential 2nd quantization of the operator of multiplication by $\chi_{[0,t)}$, that is,
\begin{eqnarray}
\Lambda (t) | \exp (f) \rangle 
=\left. \frac{d}{du}%
|\exp (e^{u\chi_{[0,t)}}f)\rangle \right| _{u=0}.
\end{eqnarray}

We have
\begin{gather}
\langle \exp \left( f\right) |\Lambda (t) |\exp \left( g\right) \rangle
 =\int_{0}^{t}f\left( s\right) ^{\ast }g\left( s \right)ds \,
\langle \exp \left( f\right) |\exp \left(
g\right) \rangle .
\end{gather}
The gauge process may be included in the quantum stochastic calculus where we now encounter the full quantum Ito table

\begin{eqnarray}
\begin{tabular}{l|llll}
$\times $ & $d \Lambda $ & $dA$ & $dA^{\ast }$ & $ dT$ \\ \hline
$d\Lambda $ & $d \Lambda$ &0 & $d A^\ast$  &0 \\
$dA^{\ast }$ &  0 & 0 & 0 & 0 \\ 
$dA$ & $dA$ &0 & $  dT$ & 0 \\ 
$dT$ & 0& 0 & 0 & 0
\end{tabular}.
\end{eqnarray}

\subsection{Unitary Dilations}
The natural question that arises now is to what extent we can realize the quantum analogues of standard concepts from classical probability theory. It turns out that the natural starting point is to think of quantum open systems and look at constructing quantum stochastic evolutions. Here one fixes a Hilbert space $\mathfrak{h}_0$, called the initial Hilbert space,  and considers quantum stochastic differential equations
\begin{eqnarray}
dU _t = \bigg( E\otimes d\Lambda_t + F \otimes dA_t^\ast + G \otimes dA_t + K \otimes dt \bigg) U_t 
\end{eqnarray}
where $E,F,G,K$ are bounded operators on the initial space. The initial condition is that $U_0$ is the identity on the tensor product of the initial space and the Fock space. This is a left-driven QSDE as the differential term in brackets appears on the left hand side. From the Ito table, one finds the necessary condition for unitarity is that the coefficients take the form
\begin{eqnarray}
E= W-I , \quad F=L , \quad G= - L^\ast W , \quad K= - \frac{1}{2} L^\ast L - iH
\label{eq:WLH}
\end{eqnarray}
where $W$ is unitary, $H$ is self-adjoint and $L$ is otherwise arbitrary.

The physical interpretation is that $H$ is the system Hamiltonian, $L$ is the coupling to noise and $S$ is a scattering operator.

The condition turns out to be sufficient \cite{HP84}. One now uses the appropriate quantum mechanism to describe the evolution, namely the Heisenberg picture $J_t (X) = U_t^\ast (X \otimes I ) U_t$. The reduced evolution where one averages over the Fock vacuum for the noise will yield a completely positive semigroup on the Banach space of operators on the initial space with Lindblad generator $\mathcal{L} (\cdot ) = \frac{1}{2} [L^\ast , \cdot ] L + \frac{1}{2} L^\ast [ \cdot , L] - i [ \cdot , H ]$.

\section{Context and Impact of Robin's Work}

Quantum mechanics developed in tandem with mathematics leading to concepts such as Hilbert space, C* and W*-operator algebras, functional integration, etc. The original problems considered dealt with isolated systems, however, it soon became clear that open systems needed to be better understood. Problems such as the radiation of photons by accelerating electrons already had the main features that would later manifest as essentials: a system, a quantum field, perturbation theory, reduced dynamics, etc. This would ultimately culminate in the theory of QED, however, this is an unreasonably detailed and cumbersome approach compared to the stochastic models that were beginning to appear elsewhere. There were several approaches to describing open quantum systems, but what was lacking was a clear concept of what a quantum stochastic process should be. The introduction of the quantum stochastic calculus changed all that.

\subsection{Interaction with Physicists}
The quantum planar Brownian motion turns out to give the description of the effective noise of a system coupled to a quantum field reservoir in the van Hove (weak coupling limit) \cite{AFL}. I played a role in this: my first\lq\lq quantum\rq\rq  paper was putting this in context \cite{AGL}, and later I managed to extend these to include scattering effects leading to the gauge process \cite{Gough05}. This meant that it was an excellent approximation for quantum optics where the higher order corrections were negligible. As a rule, physicists working in quantum optics are very happy to accept the Markov approximation; it is highly accurate and makes life simpler.

A more formal theory of quantum Brownian motion was presented in 1985 by Gardiner and Collett \cite{GC85,GZ}, one year after the Hudson-Parthasarathy paper, which built upon the input-output formalism. 

It is interesting to compare the two papers. Robin and Partha were concerned with developing a rigorous theory that built on the creation, annihilation and gauge processes, they view the quantum stochastic flows as unitary dilations of the quantum dynamical semi-groups and, as such, viewed it as a mathematical construct rather than a physical description. (Robin, in particular, was initially reluctant to view the noise model as something that could represent actual experimentally measurable objects. The talks at the 2014 Newton Institute on models and practical experiments on photon detection finally convince him of this.)

Gardiner and Collett, on the other hand, considered a physical scattering process that was made singular by means of a so-called first Markov approximation, and which gave rise to input-output relations where the inputs satisfied a singular ($\delta$-correlated) canonical commutations relations. In the latter theory, the quantum stochastic calculus emerged as a way to regularize the quantum white noise inputs. The commonality was that both papers produced a quantum Ito table and deduce the form of a unitary quantum stochastic process. It should be noted that the Gardiner-Collett Ito table does not consider the gauge process - interestingly, the Hudson-Parthasarathy paper \cite{HP84} initially came in two parts (quantum Brownian and quantum Poissonian), but an anonymous referee fortuitously advised their combination into a single paper.  

It is worth remarking that the input-output theory did not just appear in 1985. 
The use of Langevin equations for open quantum optical systems goes back at least to the work of Louisell, see \cite{Louisell} and was systematically applied to laser systems by Haken \cite{Haken}. Motivated by Haken's approach Gardiner and Collett published an important paper on squeezing \cite{GC84} which considers the Langevin equation
\begin{eqnarray}
\frac{d}{dt} a = - \frac{i}{\hbar} [ a , H_{\mathrm{sys}} ] - \frac{\gamma}{2} a + \Gamma ,
\label{eq:OU1}
\end{eqnarray}
for a cavity mode with annihilator $a$, Hamiltonian $H_{\mathrm{sys}} $, and noise $\Gamma$. They argue that the source of noise should be an external quantum white noise field $a_{\mathrm{in}}$ with $\Gamma = \sqrt{\gamma} a_{\mathrm{in}}$ and $[  a_{\mathrm{in}} (t) , a_{\mathrm{in}}^\ast (s) ] = \delta (t-s)$. Here, the input and output are related via a boundary condition $ \sqrt{\gamma} a=  a_{\mathrm{in}} + a_{\mathrm{out}}$. It was in their 1985 paper the following year that they showed that the annihilation process $A_{\mathrm{in}} (t) = \int_0^t a_{\mathrm{in}} (s) ds$ and its creator analogue form the basis for a quantum Ito theory, and that (\ref{eq:OU1}) is the quantum stochastic differential equation for $a(t) = U^\ast_t ( a \otimes I) U_t$ for an unitary process $U_t$.

The equation (\ref{eq:OU1}) would be interpreted in the Hudson-Parthasarathy theory as a quantum Ornstein-Uhlenbeck process with quantum stochastic differential equation:
\begin{eqnarray}
da =  - \frac{i}{\hbar} [ a , H_{\mathrm{sys}} ] \, dt - \frac{\gamma}{2} a ,\, dt  + \sqrt{\gamma} d A_t ,
\label{eq:OU2}
\end{eqnarray}

The unitary process $U_t$ required has coefficients, see (\ref{eq:WLH}) given by $W=I, L= \sqrt{\gamma} a$, and $ H= H_{\mathrm{sys}}$ so we may realize $a$ as the mode annihilator in the Heisenberg picture. The output noise is then given by 
\begin{eqnarray}
A_{\mathrm{out}} (t) = U_t^\ast (I \otimes A_{\mathrm{in}} (t) )  U_t .
\label{eq:IO}
\end{eqnarray}
This last identity turns out to be universal for quantum Markov models. 

As such, Gardiner and Collett's contribution was to continue the development of the theory of quantum Langevin models, but without the gauge.
In a later edition of Gardiner's book on Quantum Noise, the gauge is eventually introduced but bizarrely credited to Belavkin. 
Conversely, the concept of output processes within the Hudson-Partha formalism appeared only later through the work of Belavkin on quantum filtering and is essentially given by (\ref{eq:IO}).

Despite the clear overlaps, the two groups do not seem to have acknowledged each other's contribution. The lack of cross referencing between the two sets of authors is difficult to explain. They had different motivations, came from different traditions, and were addressing different audiences, but even this does not explain things. Robin told me that at the time he wrote to Gardiner to point out his 1984 paper, but that unfortunately seems to have been the limit of interaction.

It was left to the next generation to start putting the mathematical and the physical elements of quantum stochastic calculus together. A major impetus for this was the development of quantum control which brought in control theorists who were naturally looking for the appropriate analogue to existing stochastic control problems: given their preference for a mathematically rigorous setting, the quantum stochastic calculus was an essential element. For a derivation of quantum filtering starting from quantum stochastic models, the review paper of Bouten, Van Handel and James is highly recommended \cite{BvHJ}.

The unitary $W$ in the coefficients $(W,L,H)$ in (\ref{eq:WLH}) was re-designated as $S$ - for scattering! - in several papers on quantum control co-authored by me \cite{GJ1,GJ2}. This builds on the general quantum stochastic calculus by adding interconnection rules: it has subsequently been called the $SLH$ (pronounced \textit{essellaitch}) formalism by the physics community.

\subsection{Interaction with Mathematicians}

An interesting precursor to the quantum stochastic calculus is the paper by Robin and Ray Streater which carries the illuminating title \lq\lq Ito's formula is the chain rule with Wick ordering\rq\rq \cite{Hudson_Streater}. The paper apparently received a cool reception at the time for its rather formal and embryonic state. It does begin by realizing that the Fock space is the appropriate setting for the noise and aims to give meaning to expression of the form
\begin{eqnarray}
dM _t= F_t \, dA_t + dA^\ast_t \, G_t + H_t \, dt .
\label{eq:HS}
\end{eqnarray}
Here, the $A$ and $A^\ast$ are (what were to become) the traditional annihilation and creation processes. The coefficients $F,G,H$ are to be operator-valued processes. The paper makes the (very restrictive) ansatz that $M_t$ is a linear combination of terms of the form $f(A_t^\ast ,t) g(A_t ,t)$.
The equation (\ref{eq:HS}) has the creator differential on the left and the annihilator on the right: the goal is to build up Wick-ordered expressions. The key result is that
\begin{eqnarray}
dM_t = \frac{\partial M_t}{\partial A_t} \, dA_t + dA_t^\ast \, \frac{\partial M_t}{\partial A_t^\ast} +\frac{\partial M_t}{\partial t} \, dt .
\end{eqnarray}
Their interpretation is the following: \lq\lq provided Wick ordering is maintained, stochastic differentiation is carried out according to the chain rule.\rq\rq

The equation (\ref{eq:HS}) should be interpreted not as an Ito equation but rather as a Stratonovich (or, as Alexander Chebotarev perhaps more correctly calls it, \textit{symmetric} \cite{Chebotarev}) equation: see also \cite{JG_Strat}. Without the proper Ito theory, one misses out that the Ito correction \textit{is} the additional term due to Wick ordering: but this had to wait until 1984 and the quantum Ito product rule, the notion of quantum stochastic differential equations with adapted coefficients, etc. (Incidentally, one of the things I failed to do was to convince Robin that there was a Stratonovich version of his theory. In retrospect this is somewhat strange as the chain rule and Leibniz rule are the very things that this calculus sets out to preserve - just as in his paper with Streater!)

In 1982, there appeared several papers by Robin and Partha presenting a more rounded presentation. In the Villa Mondragone meeting there was the paper on quantum diffusions for a bose mode coupled to a (bose) Fock environment \cite{QPI} where the eventual \cite{HP84} starts to become more recognizable. There is also a different approach to quantum dilations which included Patrick Ion \cite{HIP}.

The advent of the 1984 paper with Partha, however, marked a major milestone in quantum probability. Beforehand QP was a somewhat abstract theory which had aimed to reset and generalize probability theory in the language of W*-algebras, but it was now given a powerful toolkit where concrete models could be formulated and solved. 

The issue of the Wick ordering became much more apparent through the subsequent work of Hans Maassen \cite{Maassen} on integral kernels, and later extended to include the gauge by P.-A. Meyer \cite{Meyer}.

A measure of the importance of the 1984 paper was that Quantum stochastic calculus was given its own AMS subject classification - \textbf{81S25}. Robin, in particular, singled out P.-A. Meyer's interest in the subject - leading to the monograph \cite{Meyer} - as a major coup for the subject. 

\bigskip

I am conscious that I have omitted a great deal of Robin's scientific contributions. I have said nothing about the double product integrals (work largely in collaboration with Sylvia Pulmannova), or quantum Levy area, quantum Yang-Baxter, quantum de Finetti theorems. Or his lively non-scientific interests: his love of hill walking, or his antipathy to Margaret Thatcher, going on the march against Tony Blair's War in Iraq, or his anti-Brexit cantata. 

\bigskip

\textit{Olga would like to thank all Robin’s mathematical friends for the support they gave to Robin from afar, and indeed, to his whole family, especially during the severest Covid ‘lockdown’; unfailing support, generously given and greatly appreciated, all through his final illness, up to his very last day and beyond}

\newpage

\section{Appendix}
I was fortunate to have one collaboration with Robin. Not a scientific one, but a musical one. At the end of 2018, I had just written the waltz below when Robin and Olga's Christmas card came in through the post and, being chronically remiss with sending cards, I dedicated the waltz to them \cite{NCW}.

\begin{figure}[h]
	\centering
		\includegraphics[width=0.80\textwidth]{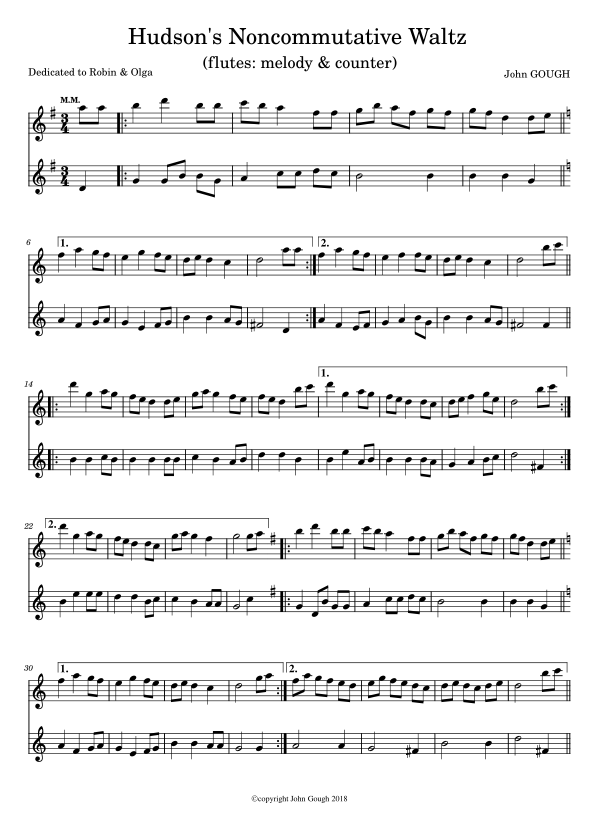}
	\label{fig:NCW1}
\end{figure}

\newpage

\begin{figure}[h]
	\centering
		\includegraphics[width=0.80\textwidth]{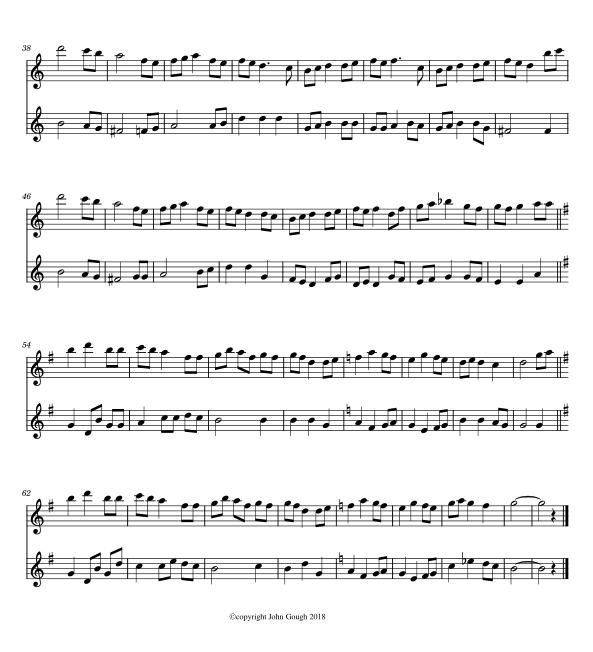}
	\label{fig:NCW2}
\end{figure}

Robin's response was: \textit{ ... maybe we could try a kind of minimally joint composition. In fact I only want to alter one note; the E in the lower part three bars from the end would become an E-flat. How well do you know your Bach chorale preludes? The very last one, \lq\lq Vor deinem Thron trit ich hiermit\rq\rq \, which I translate loosely (bearing in mind the music) as \lq\lq Before thy throne I slip in gently, hoping nobody will notice\rq\rq \, ends with a cadence which, assuming it's also in G major which I think is the case, flattens both F-sharp to F and E to E-flat. Instead of changing the mode, it gives the G major cadence extraordinarily peaceful finality. My proposal would also emphasize G major.} As per Robin's wishes, the piece was played at his funeral and memorial service- the only one not by Bach!

\bigskip

Despite the international reputation he had garnered, Robin was always self-effacing. He preferred not to claim attention for himself: unlike many of the better known researchers in quantum open systems, everyone knows exactly what his contributions were - and precisely why they were important. That, it seems, is the way he would have always liked it. Of course, despite his best efforts to discreetly slip in and out, those who knew him never failed to notice his presence. He will be missed greatly by family and friends, the QP community, and the mathematical physics community in general.

\end{document}